\documentclass[12pt]{article}
\usepackage{a4wide}
\usepackage{latexsym}
\usepackage{cite}

\usepackage{pslatex}
\usepackage{graphicx}
\usepackage[latin1]{inputenc}
\usepackage[T1]{fontenc}

\def\bq{\begin{eqnarray}}
\def\eq{\end{eqnarray}}

\begin{document}

\thispagestyle{empty}

\begin{flushright}
  MZ-TH/04-17 \\
\end{flushright}

\vspace{1.5cm}

\begin{center}
  {\Large\bf Numerical evaluation of multiple polylogarithms\\
  }
  \vspace{1cm}
  {\large Jens Vollinga and Stefan Weinzierl\\
  \vspace{1cm}
      {\small \em Institut f{\"u}r Physik, Universit{\"a}t Mainz,}\\
      {\small \em D - 55099 Mainz, Germany}\\
  } 
\end{center}

\vspace{2cm}

\begin{abstract}\noindent
  {
Multiple polylogarithms appear in analytic calculations of higher order corrections
in quantum field theory.
In this article we study the numerical evaluation of multiple polylogarithms.
We provide algorithms, which allow the evaluation for arbitrary complex arguments and without
any restriction on the weight.
We have implemented these algorithms with arbitrary precision arithmetic in C++ within the GiNaC framework.
   }
\end{abstract}

\vspace*{\fill}

\newpage

\section{Introduction}
\label{sec:intro}

In recent years multiple polylogarithms 
\cite{Goncharov,Borwein,Gehrmann:2000zt}
found the way into physics in the field of perturbative calculations
in quantum field theory.
They satisfy several useful algebraic properties, in particular they satisfy two different Hopf algebras.
Multiple polylogarithms are generalisations of
harmonic polylogarithms \cite{Vermaseren:1998uu,Remiddi:1999ew},  Nielsen polylogarithms
and the classical polylogarithms \cite{lewin:book,Nielsen} to multiple scales.
Polylogarithms have a natural grading defined by their weight.
From explicit higher order calculations it is emerging that one can express
the results of Feynman integrals in terms of multiple polylogarithms.
In dimensional regularisation we obtain in the finite terms polylogarithms up to weight $2 l$ for
an $l$-loop amplitude.
Whereas we expect harmonic polylogarithms to be sufficient to express the results 
for $2 \rightarrow 2$ scattering processes in massless quantum field theories, amplitudes with more
external particles and/or massive particles involve additional scales, which naturally leads to multiple
polylogarithms.
Since many recent results 
\cite{Bern:2000ie,Bern:2000dn,Anastasiou:2000kg,Anastasiou:2000ue,Anastasiou:2000mv,Anastasiou:2001sv,Glover:2001af,Bern:2001dg,Bern:2001df,Bern:2002tk,Garland:2001tf,Garland:2002ak,Moch:2002hm,Moch:2004pa,Vogt:2004mw,Bonciani:2003hc,Aglietti:2004tq,Bernreuther:2004ih,Korner:2004nh,Birthwright:2004kk}
of higher order calculations are expressed in terms of multiple
polylogarithms, a numerical evaluation routine for multiple polylogarithms 
is of immediate use for perturbative calculations in particle physics.
Algorithms for the numerical evaluation of multiple polylogarithms are the subject of this article.
For a few specific subclasses of multiple polylogarithms numerical evaluation methods can be found in the literature:
The numerical evaluation of Nielsen polylogarithms has been known for a long time
\cite{Kolbig:1970,Kolbig:1986qt,'tHooft:1979xw}.
A special case of multiple polylogarithms are multiple zeta values, which are obtained from the polylogarithms
if all scales are equal to one.
For this special case, efficient algorithms for the numerical evaluation have been studied by
Crandall \cite{Crandall:1998} and Borwein et al. \cite{Borwein}.
Furthermore, two recent papers provide numerical routines for harmonic polylogarithms and 
two-dimensional harmonic polylogarithms
\cite{Gehrmann:2001pz,Gehrmann:2001jv,Gehrmann:2002zr}.
However these last mentioned routines are restricted to not more than two scales and weight not higher than $4$.
Here we provide methods for the larger class of multiple polylogarithms 
without any restrictions on the weight and the number of scales.
We have implemented these algorithms
with arbitrary precision arithmetic in C++ within the GiNaC \cite{Bauer:2000cp}
framework.

This paper is organised as follows:
In Section \ref{sec:dilog} we review as an introductory example the numerical evaluation of the dilogarithm.
Section \ref{sec:def} defines the notation for multiple polylogarithms used in this paper.
Section \ref{sec:prop_polylogs} collects several useful properties of multiple polylogarithms, which will be used
to construct the algorithms for the numerical evaluation.
Section \ref{sec:num_gen} is the main part of this paper and gives the algorithms for the numerical evaluation.
In Section \ref{sec:checks} we report on the implementation in C++ within the GiNaC framework
and describe the checks that we have performed.
Finally Section \ref{sec:concl} contains our conclusions.

\section{The dilogarithm}
\label{sec:dilog}

In this short section we review as an introductory example the numerical evaluation 
of the dilogarithm \cite{'tHooft:1979xw}.
The numerical evaluation algorithm is rather simple, but does contain many ideas which will
reappear in more elaborate form in the remaining part of the paper.
The dilogarithm is defined by
\bq
\mbox{Li}_{2}(x) & = & - \int\limits_{0}^{x} dt \frac{\ln(1-t)}{t},
\eq
and has a branch cut along the positive real axis, starting at the point $ x=1$.
For $\left|x\right| \le 1$ one has the convergent power series expansion
\bq
\label{Li2power}
\mbox{Li}_{2}(x) & = & \sum\limits_{n=1}^{\infty} \frac{x^{n}}{n^{2}}.
\eq
The first step for a numerical evaluation consists in mapping an arbitrary argument $x$ into
the region, where the power series in eq. (\ref{Li2power}) converges.
This can be done with the help of the reflection identity
\bq
\label{Li2inversion}
\mbox{Li}_2(x) & = & -\mbox{Li}_2\left(\frac{1}{x}\right) -\frac{\pi^2}{6} -\frac{1}{2} \left( \ln(-x) \right)^2,
\eq
which is used to map the argument $x$, lying outside the unit circle into the unit circle.
The function $\ln(-x)$ appearing on the r.h.s. of eq. (\ref{Li2inversion}) is considered 
to be ``simpler'', e.g. it is assumed that a numerical evaluation routine for this function is known.
In addition we can shift the argument into the range
$-1 \leq \mbox{Re}(x) \leq 1/2$ with the help of
\bq
\label{Li2reflection}
\mbox{Li}_2(x) & = & -\mbox{Li}_2(1-x) + \frac{\pi^2}{6} -\ln(x) \ln(1-x).
\eq
Although one can now attempt a brute force evaluation of the power series in eq. (\ref{Li2power}), it is more
efficient to rewrite the dilogarithm as a series involving the Bernoulli numbers $B_n$:
\bq
\label{Li2Bernoulli}
\mbox{Li}_2(x) & = & \sum\limits_{i=0}^\infty \frac{B_i}{(i+1)!} z^{i+1},
\eq
with $z = - \ln(1-x)$.
The Bernoulli numbers $B_n$ are defined through the generating function
\bq
\frac{t}{e^t-1} & = & \sum\limits_{n=0}^\infty B_n \frac{t^n}{n!}.
\eq
Therefore the numerical evaluation of the dilogarithm consists 
in using eqs. (\ref{Li2inversion}) and (\ref{Li2reflection})
to map any argument $x$ into the unit circle with the additional condition $\mbox{Re}(x) \le 1/2$.
One then uses the series expansion in terms of Bernoulli numbers eq. (\ref{Li2Bernoulli}).

It occurs quite frequently in physics, that the numerical value for the dilogarithm is needed for the case,
where the variable $x$ is real except for a small imaginary part.
The small imaginary part specifies if the dilogarithm should be evaluated above or below
the cut in the case $x>1$.
In this case the evaluation is split into the evaluation for the real part and the one for the imaginary part.
The imaginary part of the dilogarithm is related to the logarithm:
\bq
\mbox{Li}_2\left( x \right) & = & 
\mbox{Re} \; \mbox{Li}_2\left( x \right) - i \ln\left(x\right) \mbox{Im} \ln \left( 1-x \right).
\eq
The imaginary part of the logarithm is given for real $x$ by
\bq
\ln\left( -x\mp i 0 \right) & = & \ln\left(\left| x \right|\right)
\mp i \pi \theta(x),
\eq
where the step function $\theta(x)$ is defined as $\theta(x)=1$ for $x>0$ and $\theta(x)=0$ otherwise
and ``$i0$'' denotes a small imaginary part.

\section{Definitions}
\label{sec:def}

\subsection{Definition of multiple polylogarithms}
\label{sec:defpolylogs}

Multiple polylogarithms are defined
by an iterated sum representation:
\bq
\mbox{Li}_{m_1,...,m_k}(x_1,...,x_k)
 & = &
\sum\limits_{i_1>i_2>\ldots>i_k>0}
     \frac{x_1^{i_1}}{{i_1}^{m_1}}\ldots \frac{x_k^{i_k}}{{i_k}^{m_k}}.
\eq
The order of the arguments of multiple polylogarithms (and of multiple zeta values)
is reversed with respect to the definitions in 
\cite{Goncharov,Moch:2001zr,Weinzierl:2002hv,Weinzierl:2003jx}.
It is also convenient to define the functions $G(z_1,...,z_k;y)$ for $z_k \neq 0$ by an
integral representation as follows \cite{Gehrmann:2000zt,Weinzierl:2003jx}:
\bq
\label{defGfct}
G(z_1,...,z_k;y) & = & \int\limits_0^y \frac{dt_1}{t_1-z_1}
 \int\limits_0^{t_1} \frac{dt_2}{t_2-z_2} ...
 \int\limits_0^{t_{k-1}} \frac{dt_k}{t_k-z_k}.
\eq
Note that in the definition of the functions $G(z_1,...,z_k;y)$
one variable is redundant due to the following scaling relation:
\bq
\label{scalingrelation}
G(z_1,...,z_k;y) & = & G(x z_1, ..., x z_k; x y)
\eq
Within the functions $G(z_1,...,z_k;y)$ one allows in addition trailing zeros (e.g $z_k = z_{k-1} = ... = z_j = 0$)
through the definitions
\bq
G(\underbrace{0,...,0}_{k};y) & = & \frac{1}{k!} \left( \ln y \right)^k,
 \nonumber \\
G(z_1,...,z_k;y) & = & \int\limits_0^y \frac{dt}{t-z_1} G(z_2,...,z_k;t).
\eq
For $z_k \neq 0$ this coincides with the original definition eq. (\ref{defGfct}).
Note that the scaling relation (\ref{scalingrelation}) does not hold for trailing zeroes. 
$G$-functions with trailing zeroes can always be transformed to $G$-functions without trailing zeroes.
An algorithm for this transformation is given in section \ref{sec:prop_polylogs}.
Therefore we will always assume that $z_k \neq 0$, unless explicitly stated otherwise.
The functions $G(z_1,...,z_k;y)$ and $\mbox{Li}_{m_1,...,m_k}(x_1,...,x_k)$
denote the same class of functions.
With the short-hand notation
\bq
G_{m_1,...,m_k}(z_1,...,z_k;y) & = &
 G(\underbrace{0,...,0}_{m_1-1},z_1,...,z_{k-1},\underbrace{0...,0}_{m_k-1},z_k;y)
\eq
the relation between the two notations is given by
\bq
\label{LiGrel}
\mbox{Li}_{m_1,...,m_k}(x_1,...,x_k) 
 & = &
 (-1)^k 
 G_{m_1,...,m_k}\left( \frac{1}{x_1}, \frac{1}{x_1 x_2}, ..., \frac{1}{x_1...x_k};1 \right).
\eq
The notation $\mbox{Li}_{m_1,...,m_k}(x_1,...,x_k)$ has a closer relation to the series representation,
whereas the notation $G(z_1,...,z_k;y)$ has a closer relation to the integral representation.
For the numerical evaluation both representations will be important.
It is further convenient to introduce the following notation for iterated integrals:
\bq
\int\limits_0^\Lambda \frac{dt}{t-a_1} \circ ... \circ \frac{dt}{t-a_n} & = & 
\int\limits_0^\Lambda \frac{dt_1}{t_1-a_1}  \times ... \times \int\limits_0^{t_{n-2}} \frac{dt_{n-1}}{t_{n-1}-a_{n-1}}\int\limits_0^{t_{n-1}} \frac{dt_n}{t_n-a_n}.
\eq
We further use the following short hand notation:
\bq
\int\limits_0^\Lambda \left( \frac{dt}{t} \circ \right)^{m} \frac{dt}{t-a}
& = & 
\int\limits_0^\Lambda 
\underbrace{\frac{dt}{t} \circ ... \frac{dt}{t}}_{m \;\mbox{\scriptsize times}} \circ \frac{dt}{t-a}.
\eq
With this notation, the integral representation of the function $G_{m_1,...,m_k}\left(z_1,z_2,...,z_k;y\right)$ 
is written as 
\bq
G_{m_1,...,m_k}\left(z_1,z_2,...,z_k;y\right)
 & = &
 \int\limits_0^y \left( \frac{dt}{t} \circ \right)^{m_1-1} \frac{dt}{t-z_1}
 \left( \frac{dt}{t} \circ \right)^{m_2-1} \frac{dt}{t-z_2}
 ...
 \left( \frac{dt}{t} \circ \right)^{m_k-1} \frac{dt}{t-z_k}.
\nonumber \\
\eq
The series expansion for $\mbox{Li}_{m_1,...,m_k}(x_1,...,x_k)$ is convergent, if
\bq
\label{condconvLi}
\left| x_1 x_2 ... x_j \right| \le 1 & & \mbox{for all} \; j \in \{1,...,k\} \; \mbox{and} \; (m_1,x_1) \neq(1,1).
\eq
Therefore the function $G_{m_1,...,m_k}\left(z_1,...,z_k;y\right)$
has a convergent series representation
if
\bq
\label{condconv}
\left| y \right| \le \left| z_j \right| \;\;\; \mbox{for all}\;j,
\eq
e.g. no element in the set $\{|z_1|,...,|z_k|,|y|\}$ is smaller than $|y|$ and in addition if $m_1=1$ we have $y/z_1 \neq 1$.
The power series expansion for the $G$-functions reads
\bq
\lefteqn{
G_{m_1,...,m_k}\left(z_1,...,z_k;y\right) = }
 \nonumber \\
 & & 
 \sum\limits_{j_1=1}^\infty
 ... 
 \sum\limits_{j_k=1}^\infty 
 \frac{1}{\left(j_1+...+j_k\right)^{m_1}} \left( \frac{y}{z_1} \right)^{j_1}
 \frac{1}{\left(j_2+...+j_k\right)^{m_2}} \left( \frac{y}{z_2} \right)^{j_2}
 ...
 \frac{1}{\left(j_k\right)^{m_k}} \left( \frac{y}{z_k} \right)^{j_k}.
\eq 
Multiple polylogarithms satisfy two Hopf algebras. The first one is referred to as ``shuffle algebra''
and is related to the integral representation.
An example for the multiplication is:
\bq
G(z_1;y) G(z_2;y) & = & G(z_1,z_2;y) + G(z_2,z_1;y).
\eq
The second algebra is often called ``stuffle algebra'' or ``quasi-shuffle algebra'' \cite{Hoffman}
and is related to the series representation.
An example for this multiplication is:
\bq
\mbox{Li}_{m_1}(x_1) \mbox{Li}_{m_2}(x_2) 
 & = &
 \mbox{Li}_{m_1,m_2}(x_1,x_2)
+\mbox{Li}_{m_2,m_1}(x_2,x_1)
+\mbox{Li}_{m_1+m_2}(x_1 x_2).
\eq

\subsubsection*{Special cases}

Multiple polylogarithms contain several specific subclasses. We list here the most 
important ones.
The notation for multiple zeta values \cite{Borwein} is:
\bq
\zeta_{m_1,...,m_k}
 & = &
\sum\limits_{i_1>i_2>\ldots>i_k>0}
     \frac{1}{{i_1}^{m_1}}\ldots \frac{1}{{i_k}^{m_k}}.
\eq
Harmonic polylogarithms \cite{Remiddi:1999ew} are denoted as
\bq
\label{harmpolylog}
H_{m_1,...,m_k}(x) & = & \mbox{Li}_{m_1,...,m_k}(x,\underbrace{1,...,1}_{k-1}).
\eq
Nielsen's polylogarithms \cite{Nielsen} are denoted as
\bq
S_{n,p}(x) & = & \mbox{Li}_{n+1,1,...,1}(x,\underbrace{1,...,1}_{p-1}).
\eq
Obviously, the class of multiple polylogarithms contains also the classical polylogarithms 
\cite{lewin:book}:
\bq
\mbox{Li}_n(x) 
 & = & 
 \sum\limits_{j=1}^\infty \frac{x^j}{j^n}.
\eq

\subsection{Definitions related to the analytic continuation}
\label{sec:analyticcont}

The multiple polylogarithms are analytic functions in $k$ complex variables.
To discuss the branch cuts it is convenient to consider the integral representation
$G_{m_1,...,m_k}\left(z_1,z_2,...,z_k;y\right)$ with $z_k \neq 0$.
\bq
G_{m_1,...,m_k}\left(z_1,z_2,...,z_k;y\right)
 & = &
 \int\limits_0^y \left( \frac{dt}{t} \circ \right)^{m_1-1} \frac{dt}{t-z_1}
 \left( \frac{dt}{t} \circ \right)^{m_2-1} \frac{dt}{t-z_2}
 ...
 \left( \frac{dt}{t} \circ \right)^{m_k-1} \frac{dt}{t-z_k}.
\nonumber \\
\eq
Using the scaling relation eq. (\ref{scalingrelation}) we can ensure that $y$ is a positive real number.
(In fact one could even require that $y=1$, but $y$ positive and real is sufficient for our purpose here.)
The $z_j$ are arbitrary complex variables.
Therefore we can deduce from the integral representation, 
that the function $G_{m_1,...,m_k}\left(z_1,z_2,...,z_k;y\right)$
develops a branch cut whenever $z_j$ is a positive real number and the integration variable $t$ exceeds $z_j$.
In most physical applications, the $z_j$ will be real numbers.
To distinguish if the integration contour runs above or below a cut, we define the abbreviations
$z_\pm$, meaning that a small positive, respectively negative imaginary part is to be added to the value of the
variable:
\bq
z_+ = z + i 0, & & z_- = z- i 0.
\eq
Close to the real axis we have
\bq
\mbox{Im}  \frac{1}{t-z\mp i0} & = & \pm \pi \frac{\partial}{\partial t} \Theta(t-z).
\eq
This formula can be used to extract the imaginary parts from the $G$-functions \cite{Moch:2002hm}.

\section{Useful properties of multiple polylogarithms}
\label{sec:prop_polylogs}

In this section we give an algorithm for the removal of trailing zeroes from $G$-functions,
discuss the transformation properties of $G$-functions $G(z_1,...,z_k; y)$ 
with respect to the argument $y$,
investigate in more detail the Bernoulli transformation and discuss the H\"older convolution.
The last two transformations can be used to speed up the series expansion.

\subsection{Trailing zeroes}
\label{sec:trailing}

In this subsection we give an algorithm to remove trailing zeroes from $G$-functions.
The method has been described for harmonic polylogarithms in \cite{Remiddi:1999ew}
and has a straightforward generalisation to multiple polylogarithms.
We say that a multiple polylogarithm of the form
\bq
G(z_1,...,z_j,\underbrace{0,...,0}_{k-j};y)
\eq
with $z_j \neq 0$
has $(k-j)$ trailing zeroes. Polylogarithms with trailing zeroes do not have a Taylor expansion in $y$, but develop
a logarithmic singularity in $(\ln y)$ around $y=0$.
In removing the trailing zeroes, one explicitly separates these logarithmic terms, such that the rest has a regular expansion
around $y=0$.
The starting point is the shuffle relation
\bq
\lefteqn{
G(0;y) G(z_1,...,z_j,\underbrace{0,...,0}_{k-j-1};y) = 
} & & \\
& &
  (k-j) G(z_1,...,z_j,\underbrace{0,...,0}_{k-j};y)
+ \sum\limits_{(s_1,...,s_{j}) = (z_1,...,z_{j-1}) \; \sqcup \!\! \sqcup \; (0)} G(s_1,...,s_j,z_{j},\underbrace{0,...,0}_{k-j-1};y).
\nonumber
\eq
$(z_1,...,z_{j-1}) \; \sqcup \!\!\! \sqcup \; (0)$ denotes the shuffle product of the string $(z_1,...,z_{j-1})$ with $(0)$.
Solving this equation for $G(z_1,...,z_j,{0,...,0};y)$ yields
\bq
\lefteqn{
G(z_1,...,z_j,\underbrace{0,...,0}_{k-j};y)
 = } & & \\
 & &
 \frac{1}{k-j} \left[
G(0;y) G(z_1,...,z_j,\underbrace{0,...,0}_{k-j-1};y) 
 -
 \sum\limits_{(s_1,...,s_{j}) = (z_1,...,z_{j-1}) \sqcup \!\! \sqcup (0)} G(s_1,...,s_j,z_{j},\underbrace{0,...,0}_{k-j-1};y)
 \right]. \nonumber 
\eq
In the first term, one logarithm has been explicitly factored out:
\bq
 G(0;y) & = & \ln y.
\eq
All remaining terms have at most $(k-j-1)$ trailing zeroes. 
Using recursion, we may therefore eliminate all trailing zeroes. 

\subsection{Transformation of the arguments}
\label{transarg}

In the following we will consider transformations of
\bq
G\left(z_1,...,z_k;f(y)\right)
\eq
with respect to the argument $y$.
For $f(y)=1-y$, $f(y)=1/(1-y)$, $f(y)=1/y$ and $f(y)=(1-y)/(1+y)$ we give algorithms, which transform
the G-function back to $G(z_1,...,z_k;y)$.
For harmonic polylogarithms the corresponding transformations have been discussed in ref. \cite{Remiddi:1999ew}.
The algorithms discussed here are straightforward extensions.

\subsubsection*{The transformation $1-y$}

The function $G(z_1,...,z_k; 1-y)$ can be expressed in terms of $G$-functions with argument $y$ and $1$
as follows:
\bq
G\left(\left.z_1\right._{\pm},...,z_k; 1-y\right)
 & = & 
 G\left(\left.z_1\right._{\pm},...,z_k; 1\right)
 +
 \int\limits_0^y \frac{dt}{t-\left(1-z_1\right)_{\mp}} G\left(z_2,...,z_k; 1-t\right).
\eq
For the second term we may use recursion.

\subsubsection*{The transformation $1/(1-y)$}

The function $G(z_1,...,z_k; 1/(1-y))$ can be expressed in terms of $G$-functions with argument $y$ and $1$
as follows: For $z_1 \neq 0$ we have
\bq
\lefteqn{
G\left(\left.z_1\right._{\pm},...,z_k; \frac{1}{1-y} \right)
 =  G\left(\left.z_1\right._{\pm},...,z_k; 1\right)
} 
\nonumber \\
 & & 
 +
 \int\limits_0^y \frac{dt}{t-\left(1-\frac{1}{z_1} \right)_{\pm}}  G\left(z_2,...,z_k; \frac{1}{1-t} \right)
 -
 \int\limits_0^y \frac{dt}{t-1_{\pm}}  G\left(z_2,...,z_k; \frac{1}{1-t} \right).
\eq
For $z_1=0$ we have
\bq
G\left(\underbrace{0,0,...,0}_{m_1-1},\left.z_2\right._{\pm},...,z_k; \frac{1}{1-y} \right)
& = & 
 G\left(\underbrace{0,0,...,0}_{m_1-1},\left.z_2\right._{\pm},...,z_k; 1\right)
\nonumber \\
& &
 -
 \int\limits_0^y \frac{dt}{t-1_{\mp}}  
 G\left(\underbrace{0,...,0}_{m_1-2},\left.z_2\right._{\pm},...,z_k; \frac{1}{1-t} \right).
\eq

\subsubsection*{The transformation $1/y$}

The function $G(z_1,...,z_k; 1/y)$ can be expressed in terms of $G$-functions with argument $y$ and $1$
as follows: For $z_1 \neq 0$ we have
\bq
\lefteqn{
G\left(\left.z_1\right._{\pm},...,z_k; \frac{1}{y} \right)
 =  G\left(\left.z_1\right._{\pm},...,z_k; 1\right)
 +
 \int\limits_0^1 \frac{dt}{t}  G\left(z_2,...,z_k; \frac{1}{t} \right)
 -
 \int\limits_0^y \frac{dt}{t}  G\left(z_2,...,z_k; \frac{1}{t} \right)
} 
\nonumber \\
 & & 
 -
 \int\limits_0^1 \frac{dt}{t-\left(\frac{1}{z_1}\right)_{\mp}}  G\left(z_2,...,z_k; \frac{1}{t} \right)
 +
 \int\limits_0^y \frac{dt}{t-\left(\frac{1}{z_1}\right)_{\mp}}  G\left(z_2,...,z_k; \frac{1}{t} \right).
\hspace*{4cm}
\eq
For $z_1=0$ we have
\bq
\lefteqn{
G\left(\underbrace{0,0,...,0}_{m_1-1},\left.z_2\right._{\pm},...,z_k; \frac{1}{y} \right)
 =  G\left(\underbrace{0,0,...,0}_{m_1-1},\left.z_2\right._{\pm},...,z_k; 1\right)
} 
\nonumber \\
 & & 
 +
 \int\limits_0^1 \frac{dt}{t}  G\left(\underbrace{0,...,0}_{m_1-2},\left.z_2\right._{\pm},...,z_k; \frac{1}{t} \right)
 -
 \int\limits_0^y \frac{dt}{t}  G\left(\underbrace{0,...,0}_{m_1-2},\left.z_2\right._{\pm},...,z_k; \frac{1}{t} \right).
\eq
Note that the $1/y$ transformation can also be obtained as a $1/(1-y)$ transformation followed by
a $1-y$ transformation.

\subsubsection*{The transformation $(1-y)/(1+y)$}

The function $G(z_1,...,z_k; (1-y)/(1+y))$ can be expressed in terms of $G$-functions with argument $y$ and $1$
as follows: For $z_1 \neq -1$ we have
\bq
\lefteqn{
G\left(\left.z_1\right._{\pm},...,z_k; \frac{1-y}{1+y} \right)
 =  G\left(\left.z_1\right._{\pm},...,z_k; 1\right)
} 
\nonumber \\
 & & 
 +
 \int\limits_0^y \frac{dt}{t-\left(\frac{1-z_1}{1+z_1}\right)_{\mp}}  G\left(z_2,...,z_k; \frac{1-t}{1+t} \right)
 -
 \int\limits_0^y \frac{dt}{t+1}  G\left(z_2,...,z_k; \frac{1-t}{1+t} \right).
\eq
For $z_1 = -1$ we have
\bq
G\left(\left.z_1\right._{\pm},...,z_k; \frac{1-y}{1+y} \right)
 & = &
 G\left(\left.z_1\right._{\pm},...,z_k; 1\right)
 -
 \int\limits_0^y \frac{dt}{t+1}  G\left(z_2,...,z_k; \frac{1-t}{1+t} \right).
\eq

\subsection{Bernoulli substitution}

In this subsection we investigate in more detail the Bernoulli transformation, 
which can be used to speed up the series expansion.
We recall the standard series representation for the dilogarithm:
\bq
\label{Li2recall}
\mbox{Li}_2(x) & = & \sum\limits_{j=1}^\infty \frac{x^j}{j^2}.
\eq
Due to eq, (\ref{Li2inversion}) and eq. (\ref{Li2reflection})
we can assume that $\left| x \right| \le 1$ and $\mbox{Re}(x) \le 1/2$.
Eq. (\ref{Li2recall}) converges therefore geometrically with $x$.
On the other hand:
\bq
\label{Li2seriesbernoulli}
\mbox{Li}_2(x) & = & 
 - \int\limits_0^x \frac{dt}{t} \ln(1-t)
 =
 \int\limits_0^{-\ln(1-x)} du \frac{u}{e^u-1}
 =
 \sum\limits_{j=0}^\infty \frac{B_j}{(j+1)!} z^{j+1},
\eq
where we used
\bq
 \frac{u}{e^u-1} & = & 
\sum\limits_{j=0}^\infty B_j \frac{u^j}{j!}
\eq
and the notation $z=-\ln(1-x)$.
For large $j$ one obtains from the formula
\bq
B_j & = & (-1)^{(j+2)/2} \frac{2 j!}{(2 \pi)^j} \zeta_j,  \;\;\; j \;\mbox{even},
\eq
that the sum converges geometrically with $z/(2\pi)$, e.g.
\bq
 \frac{B_j}{(j+1)!} z^{j+1} & = & (-1)^{(j+2)/2} \frac{2 z \zeta_j}{j+1} \left( \frac{z}{2\pi} \right)^j.
\eq
For $x$ in the range $[-1,1/2]$ we have
\bq
 \left| -\frac{\ln(1-x)}{2\pi} \right| \le \frac{\ln 2}{2\pi} < 0.1104.
\eq
Therefore we can expect an improvement in the speed of convergence 
by using eq. (\ref{Li2seriesbernoulli}) instead of eq. (\ref{Li2recall}).
We can generalise the Bernoulli transformation to the classical polylogarithms and to the harmonic polylogarithms.
For the classical polylogarithms we find
\bq
\mbox{Li}_n(x) & = & \sum\limits_{j=0}^\infty
 \frac{C_{n}(j)}{(j+1)!} \left( - \ln(1-x) \right)^{j+1},
\eq
where $C_{1}(j) = \delta_{j,0}$ and
\bq
C_{n+1}(j) & = & \sum\limits_{k=0}^j
\left( \begin{array}{c} j \\ k \\ \end{array} \right)
 \frac{B_{j-k}}{k+1} C_{n}(k).
\eq
The coefficients $C_{n}(j)$ are independent of $x$ and need therefore to be 
calculated only once.
Similar we obtain for the harmonic polylogarithms:
\bq
H_{m_1,...,m_k}(x) & = &
 \sum\limits_{j=0}^\infty
 \frac{C_{m_1,...,m_k}(j)}{(j+1)!} \left( - \ln(1-x) \right)^{j+1},
\eq
where
\bq
C_{1,m_2,...,m_k}(j) & = & \left\{ 
 \begin{array}{ll}
 0, & j=0, \\ 
 C_{m_2,...,m_k}(j-1), & j>0, \\
 \end{array}
 \right.
\eq
and 
\bq
C_{m_1+1,m_2,...,m_k}(j) & = & \sum\limits_{k=0}^j
\left( \begin{array}{c} j \\ k \\ \end{array} \right)
 \frac{B_{j-k}}{k+1} C_{m_1,m_2,...,m_k}(k).
\eq
For the classical polylogarithms $\mbox{Li}_n(x)$ the Bernoulli transformation leads to a speed-up if $n$ is not too large.
Empirically we find that for $n \ge 12$ the ``standard'' series expansion without the Bernoulli transformation is faster.
This can be related to the explicit factor $1/j^{n}$ in the series expansion
\bq
\mbox{Li}_n(x) & = & \sum\limits_{j=1}^\infty \frac{x^j}{j^n}.
\eq

\subsection{H\"older convolution}

The multiple polylogarithms satisfy the H\"older convolution \cite{Borwein}.
For $z_1 \neq 1$ and $z_w \neq 0$ this identity reads
\bq
\label{defhoelder}
G\left(z_1,...,z_w; 1 \right) 
 & = & 
 \sum\limits_{j=0}^w \left(-1\right)^j 
  G\left(1-z_j, 1-z_{j-1},...,1-z_1; 1 - \frac{1}{p} \right)
  G\left( z_{j+1},..., z_w; \frac{1}{p} \right).
 \nonumber \\
\eq
The H\"older convolution can be used to improve the rate of convergence for the series
representation of multiple polylogarithms.

\section{Numerical evaluations}
\label{sec:num_gen}

\subsection{Numerical evaluation of multiple zeta values}

Every multiple zeta value $\zeta(s_1,\ldots,s_k)$ with $s_1>1$ is finite and
can be written as a convergent sum. Therefore it seems that the numerical evaluation is
straightforward and no further transformations are needed. Yet the convergence
can be quite slow if the parameters are small or high precision is needed.
Therefore algorithms have been invented to accelerate the evaluation
\cite{Borwein:1996yq,Crandall:1995,Crandall:1998}.
The motivation often was to find numerically new relations among multiple zeta values, and to
this end the algorithms needed to deliver a very high digit accuracy, i.e.
several hundred decimal digits, efficiently in short time. Of course, low
precision evaluations benefit from these algorithms as well and since multiple zeta values with
small parameters appear in the transformations of harmonic and multiple
polylogarithms, these acceleration techniques are also important for the 
numerical evaluation of those functions.

The algorithm proposed by Crandall \cite{Crandall:1998} partitions the iterated
integral by a chosen constant $\lambda$ in such a way that the multiple zeta values can be
expressed as a finite sum over two iterated integral functions $Y$ and $Z$, for
which all integration variables are either smaller or greater than $\lambda$
respectively. Different acceleration method apply for $Y$ and $Z$. 

$Z$ can be written in almost exactly the way as $\zeta$ but has an additional
factor $e^{-\lambda n_1}$ ($n_1$ is the outermost summation index).  This factor
improves the convergence, and the greater $\lambda$ is the faster $Z$
converges. 

If $\lambda$ is chosen to be smaller than $2\pi$, $Y$ can be rewritten in terms
of Bernoulli numbers. The remaining summation can be interpreted as a
convolution and this convolution can be made fast by applying FFT
methods. The vectors to be convoluted and containing Bernoulli numbers or factors
of gamma functions have to have a certain length $L$, which becomes a parameter
of the algorithm like $\lambda$.  The performance of the evaluation depends on
the values of $\lambda$ and $L$, but quite generally one can find that
Crandall's algorithm performs best for high precision and a small number of
$\zeta$ parameters.

Another method \cite{Borwein} uses the
H\"older convolution. As a generalisation of duality a given $\zeta_{m_1,\ldots,m_k}$ 
is rewritten as a
finite sum over functions 
\bq
\mbox{G}_{m_1,\ldots,m_k}(1,\ldots,1;1/p) \;\;\;\mbox{and}\;\;\; \mbox{G}_{m_1,\ldots,m_k}(1,\ldots,1;1/q).
\eq
For $p,q > 1$ they have a significantly better convergence than $\zeta_{m_1,\ldots,m_k}$.
The $p$ and $p$ must satisfy the H\"older condition $1/p + 1/q = 1$
and can be chosen for best convergence as $p=q=2$.

The acceleration is significant and exceeds that of Crandall's algorithm for
low precision evaluations.  Both algorithms can play hand in hand to satisfy
different demands. For high precision evaluations Crandall's algorithm should
be chosen, for the other cases H\"older convolution is the best choice.

\subsection{Numerical evaluation of harmonic polylogarithms}

Harmonic polylogarithms have been introduced by Remiddi and Vermaseren \cite{Remiddi:1999ew} and
their numerical evaluation has been worked out by Gehrmann and Remiddi \cite{Gehrmann:2001pz}.
Harmonic polylogarithms have a convergent series representation if $|x| < 1$ and no trailing zeroes
are present.
The rightmost indices are allowed to be zero, in which case the harmonic polylogarithm
does not have a series expansion around zero in $x$, but develops a logarithmic singularity
proportional to $\ln x$. 
To cope with this trailing zeros scenario, one uses recursively the
shuffle algebra to extract those zeros in form of ordinary logarithms.  
It remains to transform an arbitrary argument $x$ into the domain $|x| < 1$.
To this aim one uses the transformations $1/x$ and $(1-x)/(1+x)$ from sect. \ref{transarg}.
Here we differ slightly from the implementation by Gehrmann and Remiddi:
These authors provide a numerical evaluation routine for harmonic polylogarithms
up to weight 4. They use for $x \in [\sqrt{2}-1,1]$ the $(1-x)/(1+x)$ transformation to improve
the convergence of the resulting series.
We provide a numerical evaluation routine for arbitrary weights and therefore have to determine the
transformation formula at run time.
We observed that this transformation by itself is quite expensive. This can outweigh the
potential gain from the series acceleration.
It is therefore more economical to apply this transformation only for $x$ near one.
The starting point is determined empirically. 

In addition, multiple zeta values appear in the transformed expressions. If some indices are negative,
alternating multiple zeta values occur.  The speed of the numerical evaluation of multiple zeta values also has a
major impact on evaluation speed of the harmonic polylogarithms.

\subsection{Numerical evaluation of multiple polylogarithms}

Multiple polylogarithms can be evaluated with the help of the series expansion
in the domain where this expansion is convergent.
This is the case if the arguments satisfy eq. (\ref{condconvLi})
or equivalently eq. (\ref{condconv}).
We first show
that any combination of arguments can be transformed 
into this domain.
This transformation is based on the integral representation and is therefore 
most naturally expressed in terms of the $G$-functions.
We consider the function
\bq
\label{generalcaseordering}
G_{m_1,...,m_k}\left(z_1,...,z_{j-1},s,z_{j+1},...,z_k;y\right),
\eq
with the condition that $|s|$ is the smallest element in the set
$\{|z_1|,...,|z_{j-1}|,|s|,|z_{j+1}|,...,|z_k|,|y|\}$.
The algorithm goes by induction 
and introduces the more general structure
\bq
\label{generalstructure}
  \int\limits_0^{y_1} \frac{ds_1}{s_1-b_1} ...
  ... \int\limits_0^{s_{r-1}}  \frac{ds_r}{s_r-b_r}  G(a_1,...,s_r,...,a_w;y_2),
\eq
where $|y_1|$ is the smallest element in the set
$\{|y_1|, |b_1'|, ..., |b_r'|, |a_1'|, ..., |a_w'|, |y_2| \}$.
The prime indicates that only the non-zero elements of $a_i$ and $b_j$ are considered.
If the integrals over $s_1$ to $s_r$ are absent, we recover the original $G$-function in eq. (\ref{generalcaseordering}).
Since we can always remove trailing zeroes with the help of the algorithm in section \ref{sec:trailing}, we can assume that
$a_w \neq 0$.
We first consider the case where the $G$-function is of depth one, e.g.
\bq
  \int\limits_0^{y_1} \frac{ds_1}{s_1-b_1} ...
  ... \int\limits_0^{s_{r-1}}  \frac{ds_r}{s_r-b_r}  G(\underbrace{0,...,0}_{m-1},s_r;y_2)
 & = &
  \int\limits_0^{y_1} \frac{ds_1}{s_1-b_1} ...
  ... \int\limits_0^{s_{r-1}}  \frac{ds_r}{s_r-b_r}  G_m(s_r;y_2),
\eq
and show that we can relate the function $G_m(s_r;y_2)$ 
to $G_m\left(y_2;s_r\right)$, powers of $\ln(s_r)$ and functions, which
do not depend on $s_r$.
For $m=1$ we have
\bq
G_1\left(\left.s_r\right._{\pm};y_2\right) & = & 
 G_1\left(\left.y_2\right._{\mp};s_r\right) - G(0;s_r) + \ln\left(-\left.y_2\right._{\mp}\right).
\eq
For $m \ge 2$ one can use the transformation $1/y$ from sect. \ref{transarg} and one obtains:
\bq
\label{oneoverxforclasspolylog}
G_m\left(\left.s_r\right._{\pm};y_2\right) & = &
 -\zeta_m + \int\limits_0^{y_2} \frac{dt}{t} G_{m-1}\left(\left.t\right._{\pm};y_2\right)
          - \int\limits_0^{s_r} \frac{dt}{t} G_{m-1}\left(\left.t\right._{\pm};y_2\right).
\eq
One sees that the first and second term in eq. (\ref{oneoverxforclasspolylog}) yield functions
independent of $s_r$.
The third term has a reduced weight and we may therefore use recursion.
This completes the discussion for $G_m\left(s_r;y_2\right)$. 
We now turn to the general case with a $G$-function of depth greater than one in eq. (\ref{generalstructure}). 
Here we first consider the sub-case, that $s_r$ appears in the last place in the parameter list and $(m-1)$ zeroes precede $s_r$, e.g.
\bq
  \int\limits_0^{y_1} \frac{ds_1}{s_1-b_1} ...
  ... \int\limits_0^{s_{r-1}}  \frac{ds_r}{s_r-b_r}  G(a_1,...,a_{k},\underbrace{0,...,0}_{m-1},s_r;y_2),
\eq
Since we assumed that the $G$-function has a depth greater than one, we may conclude that $a_k \neq 0$.
Here we use the shuffle relation to relate this case to the case where $s_r$ does not appear in the last place:
\bq
\label{shuffleG}
G(a_1,...,a_{k},\underbrace{0,...,0}_{m-1},s_r;y_2)
 & = & 
 G(a_1,...,a_{k};y_2) G(\underbrace{0,...,0}_{m-1},s_r;y_2)
 - \sum\limits_{shuffles'} G(\alpha_1,...,\alpha_{k+m};y_2),
 \nonumber \\
\eq
where the sum runs over all shuffles of $(a_1,...,a_{k})$ with $(0,...,0,s_r)$ and the prime indicates that
$(\alpha_1,...,\alpha_{k+m}) = (a_1,...,a_{k},0,...,0,s_r)$ is to be excluded from this sum.
In the first term on the r.h.s of eq. (\ref{shuffleG}) the factor $G(a_1,...,a_{k};y_2)$
is independent of $s_r$ , whereas the second factor $G(0,...,0,s_r;y_2)$ is of depth one
and can be treated with the methods discussed above.
The terms corresponding to the sum over the shuffles in eq. (\ref{shuffleG}) have either $s_r$ not appearing in the last place
in the parameter list or a reduced number of zeroes preceding $s_r$. In the last case we may use recursion to remove $s_r$ from
the last place in the parameter list.
It remains to discuss the case, where the $G$-function has depth greater than one and $s_r$ does not appear in the last
place in the parameter list, e.g.
\bq
  \int\limits_0^{y_1} \frac{ds_1}{s_1-b_1} ...
  ... \int\limits_0^{s_{r-1}}  \frac{ds_r}{s_r-b_r}  G(a_1,...,a_{i-1},s_r,a_{i+1},...,a_w;y_2),
\eq
with $a_w \neq 0$.
Obviously, we have
\bq
\lefteqn{
 G\left(a_1,...,a_{i-1},s_r,a_{i+1},...,a_{w};y_2\right)
 = } & & \\
& &
  G\left(a_1,...,a_{i-1},0,a_{i+1},...,a_{w};y_2\right)
  + \int\limits_0^{s_r} ds_{r+1} \frac{\partial}{\partial s_{r+1}}  
    G\left(a_1,...,a_{i-1},s_{r+1},a_{i+1},...,a_{w};y_2\right). \nonumber 
\eq
The first term $G\left(a_1,...,a_{i-1},0,a_{i+1},...,a_{w};y_2\right)$ does no longer depend on $s_r$ and has a reduced depth.
For the second term we first write out the integral representation of the $G$-function. We then use
\bq
\frac{\partial}{\partial s} \frac{1}{t-s} 
 & = & 
 - \frac{\partial}{\partial t} \frac{1}{t-s},
\eq
followed by partial integration in $t$ and finally partial fraction decomposition
according to
\bq
 \frac{1}{t-\alpha} \frac{1}{t-s}
 & = & 
 \frac{1}{s-\alpha} \left( \frac{1}{t-s} - \frac{1}{t-\alpha} \right).
\eq
If $s_r$ is not in the first place of the parameter list, we obtain
\bq
\lefteqn{
\int\limits_0^{s_r} ds_{r+1} \frac{\partial}{\partial s_{r+1}}  
    G\left(a_1,...,a_{i-1},s_{r+1},a_{i+1},...,a_{w};y_2\right)
} \nonumber \\
 & = &
 - \int\limits_0^{s_r} \frac{ds_{r+1}}{s_{r+1}-a_{i-1}}
   G\left(a_1,...,a_{i-2},s_{r+1},a_{i+1},...,a_w;y_2\right)
 \nonumber \\
 & &
 + \int\limits_0^{s_r} \frac{ds_{r+1}}{s_{r+1}-a_{i-1}}
   G\left(a_1,...,a_{i-2},a_{i-1},a_{i+1},...,a_w;y_2\right)
 \nonumber \\
 & &
 + \int\limits_0^{s_r} \frac{ds_{r+1}}{s_{r+1}-a_{i+1}}
   G\left(a_1,...,a_{i-1},s_{r+1},a_{i+2},...,a_w;y_2\right)
 \nonumber \\
 & &
 - \int\limits_0^{s_r} \frac{ds_{r+1}}{s_{r+1}-a_{i+1}}
   G\left(a_1,...,a_{i-1},a_{i+1},a_{i+2},...,a_w;y_2\right).
\eq
Each $G$-function has a weight reduced by one unit and we may use recursion.
If $s_r$ appears in the first place we have the following special case:
\bq
\lefteqn{
\int\limits_0^{s_r} ds_{r+1} \frac{\partial}{\partial s_{r+1}}  
    G\left(s_{r+1},a_{i+1},...,a_{w};y_2\right)
 = 
 \int\limits_0^{s_r} \frac{ds_{r+1}}{s_{r+1}-y_2}
   G\left(a_{i+1},...,a_w;y_2\right)
} & & \nonumber \\
 & &
 + \int\limits_0^{s_r} \frac{ds_{r+1}}{s_{r+1}-a_{i+1}}
   G\left(s_{r+1},a_{i+2},...,a_w;y_2\right)
 - \int\limits_0^{s_r} \frac{ds_{r+1}}{s_{r+1}-a_{i+1}}
   G\left(a_{i+1},a_{i+2},...,a_w;y_2\right).
\eq
There is however a subtlety: If $\alpha_{i-1}$ or $\alpha_{i+1}$ are zero, the algorithm generates
terms of the form
\bq
 \int\limits_0^y \frac{ds}{s} F(s) 
- \int\limits_0^y \frac{ds}{s} F(0).
\eq
Although the sum of these two terms is finite, individual pieces diverge at $s=0$.
We regularise the individual contributions with a lower cut-off $\lambda$:
\bq
 \int\limits_\lambda^y \frac{ds}{s} F(s) 
- \int\limits_\lambda^y \frac{ds}{s} F(0).
\eq
In individual contributions we therefore obtain at the end of the day powers of $\ln\lambda$ from integrals of the form
\bq
\int\limits_\lambda^y \frac{ds_1}{s_1} \int\limits_\lambda^{s_1} \frac{ds_2}{s_2} 
 & = & \frac{1}{2} \ln^2 y - \ln y \ln \lambda + \frac{1}{2} \ln^2 \lambda.
\eq
In the final result, all powers of $\ln \lambda$ cancel, and we are left with $G$-functions with trailing zeros.
These are then converted by standard algorithms to $G$-functions without trailing zeros.
The $G$-functions without trailing zeros can then be evaluated numerically by their power series expansion.

In addition, the algorithms may introduce in intermediate steps $G$-functions with leading
ones, e.g $G(1,...,z_k;1)$.
These functions are divergent, but the divergence can be factorised and expressed in terms
of the basic divergence $G(1;1)$. 
The algorithm is very similar to the one for the extraction of trailing zeroes.
In the end all divergences cancel.

\subsubsection*{Acceleration of the convergent series}

The $G$-function $G_{m_1,...,m_k}(z_1,...,z_k;y)$ has a convergent power series expansion if the conditions
in eq. (\ref{condconv}) are met. This does not necessarily imply, that the convergence is sufficiently fast, such
that the power series expansion can be used in a straightforward way.
In particular, if $z_1$ is close to $y$ the convergence is rather poor.
In this paragraph we consider methods to improve the convergence.
We can assume that no trailing zeroes are present ($z_k \neq 0$), therefore we can normalise $y$ to one.
Convergence implies then, that we have $|z_j| \ge 1$ and $(z_1,m_1) \neq (1,1)$.
If some $z_j$ is close to the unit circle, say,
\bq
\label{condhoelder}
1 \le \left| z_j \right| \le 2,
\eq
we use the H\"older convolution eq. (\ref{defhoelder}) with $p=2$ to rewrite the $G$-functions as
\bq
\label{localhoelder}
\lefteqn{
G\left(z_1,...,z_w; 1 \right) 
 =  
 G\left(2 z_1,..., 2 z_w; 1 \right)
 + (-1)^w G\left( 2(1-z_w), 2(1-z_{w-1}),..., 2(1-z_1); 1 \right)
 } & &
 \nonumber \\
 & & 
 + \sum\limits_{j=1}^{w-1} \left(-1\right)^j 
  G\left( 2(1-z_j), 2(1-z_{j-1}),..., 2(1-z_1); 1 \right)
  G\left( 2 z_{j+1},..., 2 z_w; 1 \right).
\eq
Here, we normalised the r.h.s to one and explicitly wrote the first and last term of the sum.
We observe, that 
the first term
$G\left(2 z_1,..., 2 z_w; 1 \right)$ has all arguments outside $\left| 2 z_j \right| \ge 2$.
This term has therefore a better convergence.
Let us now turn to the second term in eq, (\ref{localhoelder}).
If some $z_j$ lies within $\left|z_j-1\right| < 1/2$, the H\"older convolution transforms the arguments out of the
region of convergence. In this case, we repeat the steps above, e.g. transformation into the region of convergence,
followed by a H\"older convolution, if necessary.
While this is a rather simple recipe to implement into a computer program, it is rather tricky to proof that this procedure does
not lead to an infinite recursion, and besides that, does indeed lead to an improvement in the convergence.
For the proof we have to understand how the algorithms for the transformation into the region of convergence act
on the arguments of a $G$-function with length $w$.
In particular we have to understand how in the result the $G$-functions of length $w$ are related to the original
$G$-function. Products of $G$-functions of lower length are ``simpler'' and not relevant for the argument here.
We observe, that this algorithm for the $G$-function $G\left(z_1,...,z_w;y\right)$ substitutes $y$ by the element
with the smallest non-zero modulus from the set $\{ \left|z_1\right|, ..., \left|z_w\right|, \left|y\right| \}$,
permutes the remaining elements into an order, which is of no relevance here and possibly substitutes
some non-zero elements by zero.
The essential point is, that it does not introduce any non-trivial new arguments (e.g. new non-zero arguments).

For the H\"older convolution we are concerned with the second term of eq. (\ref{localhoelder})
\bq
\label{Hoelderterm}
G\left( 2(1-z_w), 2(1-z_{w-1}),..., 2(1-z_1); 1 \right).
\eq
The first term $G\left(2 z_1,..., 2 z_w; 1 \right)$ never transforms the arguments into the
non-convergent region and has, as we have seen, a better convergence.
The terms in the sum of eq. (\ref{localhoelder}) have a reduced length, 
and by induction we can assume that suitable methods to improve the convergence
exist for those terms.
For the second term of eq. (\ref{localhoelder}) we have to discuss the transformation
$z \rightarrow 2(1-z)$.
We divide the arguments $z_j$ into different classes:
\begin{description}
\item{-} class A: $\left|z\right|>2$. These will map under the transformation
$z \rightarrow 2(1-z)$ again to $\left|z\right|>2$. Actually, they transform to $\left|z-2\right|>4$, but this region
is included in the previous one.
\item{-} class B: $1 \le \left|z\right| \le 2$ and $\left|z-1\right| > 1$. 
These will map under the transformation $z \rightarrow 2(1-z)$ to $\left|z\right| > 2$ (and
necessarily $\left|z-1\right| > 1$).
Therefore class B is mapped to class A.
\item{-} class C: $1 \le \left|z\right| \le 2$ and $1/2 \le \left|z-1\right| \le 1$.
These will map under the transformation $z \rightarrow 2(1-z)$ to $1 \le \left|z\right| \le 2$
and $\left|z-1\right| > 1$.
Therefore class C is mapped to class B.
\item{-} class D: $1 \le \left|z\right| \le 2$ and $\left|z-1\right| < 1/2$.
These will map under the transformation $z \rightarrow 2(1-z)$ to $\left|z\right| < 1$ and 
$\left|z-1\right| > 1$.
Class D is mapped to the non-convergent region $\left| z \right| < 1$.
\end{description}
Let us first assume that all $z_j$'s are from class A and B. Then after the H\"older convolution, all arguments
satisfy $\left| z \right| > 2$, which ensures a fast convergence.
Secondly, we assume that all $z_j$'s are from the classes A, B and C. The the H\"older convolution will
generate $G$-functions with arguments from the classes A and B alone. One subsequent H\"older convolution
on those $G$-functions, which contain arguments from class B will again lead to
$\left| z \right| > 2$ for all $z_j$.
For the last case we have to consider $G$-functions, where some arguments are from class D.
Then we obtain arguments with $\left| z \right| < 1$ and it is necessary to re-use the transformation into the
region of convergence.
Let $n_{CD}$ be the number of arguments $z_j$ from the classes C and D from the 
original $G$-function.
The second term in eq. (\ref{localhoelder}) equals eq. (\ref{Hoelderterm}).
Let $z_{min}$ be the argument such that $2|1-z_{min}|$ is the smallest in the set
\bq
\left\{ 2 \left|1-z_w \right|, ..., 2 \left|1-z_1 \right|, 1 \right\}.
\eq
Since at least one argument is from class D, we have $2|1-z_{min}| < 1$.
The algorithm for the transformation into the convergent region introduces then
$G$-functions of the form
\bq
\label{anotherGfunction}
G\left( \frac{1-z_{\sigma_{w}}}{1-z_{min}}, ..., \frac{1-z_{\sigma_{j+1}}}{1-z_{min}},
        \frac{1}{2(1-z_{min})}, \frac{1-z_{\sigma_{j-1}}}{1-z_{min}}, ...,
        \frac{1-z_{\sigma_{1}}}{1-z_{min}}; 1 \right).
\eq
Here, $\sigma$ is permutation of the original arguments. The argument $1/2/(1-z_{min})$ results from permutating
the original $y=1$ into the argument list of the $z$'s.
We note that this argument satisfies
\bq
\label{argyeqone}
\left| \frac{1}{2(1-z_{min})} -1 \right| > 1,
\eq
and lies therefore always outside region C and D.
This is most easily seen by using polar coordinates for $2(1-z_{min})$.
Furthermore, if an arbitrary argument $z$ is from class A or B, it satisfies $|z-1| > 1$.
Therefore
\bq
\left| z - z_{min} \right| > \frac{1}{2},
\eq
and it follows that
\bq
\left| \frac{1-z}{1-z_{min}} - 1 \right| > 1.
\eq
Therefore, arguments from classes A and B will remain in these classes.
It remains to consider the case $z=0$ and to show it does not re-introduce arguments in the classes C or D.
Again, we can show that
\bq
\left| \frac{1}{1-z_{min}} -1 \right| > 1.
\eq
In summary, the $G$-function in eq. (\ref{anotherGfunction}) has the number $n_{CD}$ reduced by at least one
(due to eq. \ref{argyeqone}).
Therefore the algorithm will successively remove the arguments from classes C and D and terminate.
This completes the proof.

In practice, there is again a trade-off between the gain in acceleration and the cost involved for the H\"older convolution.
We therefore apply the H\"older convolution only if some $z_j$ satisfies
\bq
1 \le \left| z_j \right| \le \lambda,
\eq
with $\lambda < 2$. A typical value is $\lambda = 0.01$.

\section{Implementation and checks}
\label{sec:checks}

The numerical evaluations have been implemented as part of GiNaC \cite{Bauer:2000cp}, a C++
library for computer algebra ({\tt http://www.ginac.de}). 
GiNaC enables symbolic algebraic manipulations within the C++ programming language.
Like FORM \cite{Vermaseren:2000nd}, it was developed within the high-energy physics community.
GiNaC allows numerics in arbitrary precision.

The following functions have been added to GiNaC. On the left are the names and
the mathematical notations, on the right are the names within GiNaC.

\begin{center}
\begin{tabular}{lll}
Name & Math. notation & GiNaC notation \\
\hline
Classical polylogarithm & $\quad \mbox{Li}_n(x)$ & {\tt Li(n,x)} \\
Nielsen polylogarithm & $\quad \mbox{S}_{n,p}(x)$ & {\tt S(n,p,x)} \\
Harmonic polylogarithm & $\quad \mbox{H}_{\vec{m}}(x)$ & {\tt H(\{m1,\ldots,mk\},x)} \\
Multiple polylogarithm & $\quad \mbox{Li}_{\vec{m}}(\vec{x})$ & {\tt Li(\{m1,\ldots,mk\},\{x1,\ldots,xk\})} \\
 & $\quad \mbox{G}(\vec{\alpha},y)$ & {\tt G(\{a1,\ldots,ak\},y)} \\
 & & {\tt G(\{a1,\ldots,ak\},\{d1,\ldots,dk\},y)} \\
Multiple zeta value & $\quad \zeta(\vec{m})$ & {\tt zeta(\{m1,\ldots,mk\})} \\
Alternating MZV & $\quad \zeta(\vec{m},\vec{\sigma})$ & {\tt zeta(\{m1,\ldots,mk\},\{s1,\ldots,sk\})}
\end{tabular}
\end{center}

While $x$ and the entries of $\vec{x}$ can be arbitrary complex numbers, $n$, $p$ and the entries of $\vec{m}$ 
must be positive integers. In the case of the harmonic polylogarithm $\vec{m}$ may also contain negative
integers. 
$\vec{\alpha}$ is the expanded parameter string for the $G$-functions.
The $d_i$'s are positive or negative numbers indicating the signs
of a small imaginary part of $\alpha_i$.
The $s_i$'s are positive or negative numbers indicating 
which corresponding sum in the
definition of the multiple zeta values is alternating. 

The algorithms used can be found in \cite{Kolbig:1970,Kolbig:1986qt} for classical polylogarithms and Nielsen
polylogarithms, in \cite{Remiddi:1999ew,Gehrmann:2001pz} for harmonic polylogarithms,
in \cite{Crandall:1998,Borwein} for multiple zeta values and in our
text above for multiple polylogarithms. The transformations are performed
algebraically so there are no restrictions on the size or length of the
parameters. The summation is accelerated by the use of Bernoulli numbers in the
case of classical and Nielsen polylogarithms. Since GiNaC has arbitrary precision
numerics and the precision can be changed at every time this is the only
feasible acceleration technique.  Evaluation of multiple zeta values either goes by Crandall's
algorithm or by H\"older convolution depending on the precision and parameters.
The switching point has been determined empirically. For alternating multiple zeta values
H\"older convolution is always used.

For the sign of the imaginary part we adopted a widely used convention for mathematical software:
{\em implementations shall map a cut so the function is continuous as the cut is approached coming
around the finite endpoint of the cut in a counter clockwise direction} \cite{C99standard}.  
With this convention the cuts on the positive real axis are 
continuous to the lower complex half-plane.
This convention also ensures consistency among the
various polylogarithms including the ordinary logarithm.
We note that the implementation of harmonic polylogarithms by Gehrmann and Remiddi \cite{Gehrmann:2001pz}
uses the opposite sign convention.
Gehrmann and Remiddi
give the argument of H a positive imaginary part $x+i\epsilon$.
This implies that cuts on the positive
real axis are continuous to the upper complex half-plane.

To ensure the correctness of the implementation many checks have been performed.
First, one can compare the numerical evaluation to known special values
of the function, e.g. $\mbox{Li}_n$ with parameters $x=0, \frac{1}{2} \mbox{ or } 1$,
or $\zeta(2n)$ with integer $n$. In the same sense but with less rigour of
course, one can compare to other well proven software implementations for a
quick reassurance. Second, the transformations can be checked for internal
consistency if they overlap. Hence the $1-x$ transformation can always be be
verified against the pure series. If the $(1-x)/(1+x)$ transformation is done
as it is with the harmonic polylogarithm the $1/x$ transformation overlaps and can be
tested as well. Checking identities between different functions is
the third method. Since all discussed functions merely are special cases of the general
multiple polylogarithm, there exists a hierarchy of identities between them,
e.g.
\[
\mbox{S}_{n,1}(x) = \mbox{Li}_{n+1}(x) \quad \mbox{or} \quad \mbox{H}_{3,1,1}(x) = \mbox{S}_{2,3}(x).
\]
In addition, limits can be examined, for example
\[
\mbox{Li}_{2,3,4}(x,a,1) \rightarrow \mbox{H}_{2,3,4}(x) \quad\mbox{as}\quad a\rightarrow 1.
\]
The checks listed above are sufficient to assure the correctness of the implementation for the simpler
functions.
Yet the more general functions of multiple polylogarithms and multiple zeta values 
still have areas of the parameter space that have not been tested with the methods so far.
Here, the shuffle and
quasi-shuffle identities are a very important tool to fill this gap. Identities
like 
\[
\zeta(4,3) = 17\zeta(7) - 10\zeta(2)\zeta(5)
\]
or
\[
\mbox{Li}_{2}(x) \mbox{Li}_{5}(y) = \mbox{Li}_{2,5}(x,y) + \mbox{Li}_{5,2}(y,x) + \mbox{Li}_{7}(xy)
\]
are examples for such identities.  
Shuffle and quasi-shuffle relations similar to those can be used to expose
problems within the implementation for arbitrary parameter sets.

The checks mentioned above have been exercised not only once, but most of them
have been added as automatic checks to the GiNaC library itself. With the
source code at hand the test suite of GiNaC can be run anytime. This helps to
detect new errors quickly that might make their way into the code in the course
of future extensions or optimisations.

The standard use of GiNaC is through its C++ interface.
A small example program is listed below:
\begin{verbatim}
#include <iostream>
#include <ginac/ginac.h>

int main()
{
  using namespace std;
  using namespace GiNaC;

  ex x1 = numeric(8,3);
  ex x2 = numeric(1,5);

  cout << "Li_{1,1}(8/3,1/5)  = "
       << Li( lst(1,1), lst(x1,x2) ).evalf() << endl;

  return 0;
}
\end{verbatim}
If this program is compiled and linked with the GiNaC-library, it will print out:
\begin{verbatim}
Li_{1,1}(8/3,1/5)  = -0.8205920210842043836-0.70102614150465842094*I
\end{verbatim}
Alternatively, GiNaC offers also 
a small interactive shell called {\it ginsh}, which allows to
try and use GiNaC's features directly as in the following examples:
\begin{verbatim}
> Li(2,1);
1/6*Pi^2
> S(2,3,4.5);
-1.5214058021507574768+1.7013776892289268546*I
> Li({2,2,1},{3.0,2.0,0.2});
-0.7890678826631402472+0.5791683703217281085*I
> Digits=40;
40
> H({2,-1,3},8.7);
-5.65207410697321998445159060623787475178342968036-1.05486293307539105482
5025378324573142440702785858*I
\end{verbatim}
Here the user input is done at the prompt {\tt >} and the result is given on
the next line.

\section{Conclusions}
\label{sec:concl}

In this paper we reported on numerical evaluation methods for multiple polylogarithms.
These functions occur in higher loop calculations in quantum field theory.
We provided algorithms, which allow the evaluation of these functions for arbitrary complex
arguments without any restriction on the weight of the function.
The functions can be evaluated in C++ to arbitrary precision within the GiNaC framework.
Subclasses of multiple polylogarithms are the classical polylogarithms, the Nielsen polylogarithms, the
harmonic polylogarithms and the multiple zeta values.
For these subclasses we have implemented specialised algorithms.
All routines are integrated in the computer algebra package GiNaC from version 1.3 onwards
and can be obtained by downloading this library \cite{Bauer:2000cp}.


\end{document}